          [1999/12/02 1.01 (PWD)]
\documentclass[a4paper,twocolumn]{esapub} 

\usepackage{natbib,epsf}
\title{SuperAGILE: the X-ray Monitor of the AGILE gamma-ray Mission}
\author[1]{M. Feroci}
\author[1]{E. Costa}
\author[1]{E. Del Monte}
\author[1]{I. Lapshov}
\author[1]{\\ M. Mastropietro}
\author[2]{E. Morelli}
\author[3]{M. Rapisarda}
\author[1]{A. Rubini}
\author[1]{P. Soffitta}
\author[4]{\\ G. Barbiellini}
\author[4]{F. Longo} 
\author[4]{M. Prest}
\author[4]{E. Vallazza}
\author[5]{\\ A. Argan}
\author[5]{S. Mereghetti}
\author[5]{M. Tavani}
\author[5]{S. Vercellone}
\author[6]{A. Morselli}

\affil[1]{Istituto di Astrofisica Spaziale, CNR, Rome, Italy }
\affil[2]{Istituto TESRE, CNR, Bologna, Italy}
\affil[3]{ENEA - Frascati, Italy}
\affil[4]{INFN - Sezione di Trieste, Italy}
\affil[5]{Istituto di Fisica Cosmica "G. Occhialini", CNR, Milan, Italy}
\affil[6]{Univ. Roma "Tor Vergata" and INFN - Sezione di Roma 2, Italy}

\begin{document}


\maketitle

\begin{abstract}
SuperAGILE is the X-ray stage of the AGILE gamma-ray mission. 
It is devoted to monitor X-ray
(10-40 keV) sources with a sensitivity better than 10 mCrabs in 50 ks and
to detect X-ray transients in a field of view of 1.8 sr, well matched to that
of the gamma-ray tracker, with few arc-minutes position resolution and
better than 5 $\mu$s timing resolution.
SuperAGILE is designed to exploit one additional layer of four Si
microstrip detectors placed on top
of the AGILE tracker, and a system of four mutually orthogonal 
one-dimensional coded masks to encode the X-ray sky. 
The total geometric area is 1444 cm$^{2}$.
Low noise electronics
based on ASIC technology composes the front-end read out.
We present here the instrumental and astrophysical performances of
SuperAGILE as derived by analytical calculations, Monte Carlo
simulations and experimental tests on a prototype of the silicon microstrip
detector and front-end electronics.
\end{abstract}

\section{Introduction}

AGILE (``Astrorivelatore Gamma ad Immagini LEggero'',  
Tavani et al. 2000 and Barbiellini et al. 2000) 
is the first mission of the Program of Small Missions 
of the Italian Space Agency (ASI). The main goal of AGILE 
is to monitor the gamma-ray sky in the energy range between 
30 MeV and 50 GeV, with a large field of view ($\sim$3 sr), good
sensitivity, good angular resolution and good timing (dead time lower
than 100 $\mu$s for the gamma-ray detector). 

AGILE is scheduled for launch by the beginning of 2003 in an equatorial
$\sim$100 minutes orbit, for a nominal lifetime of 2 years. 
It will use the ASI base in Malindi as a ground station. 
The satellite mass will be about 200 kg ($\sim$65 kg of payload) and
the power available to the payload is about 65 W. The scientific telemetry
will be able to transmit approximately 300 Mbit of data to ground
at each passage over the Malindi ground station.

The AGILE payload (Mereghetti et al. 2000)
is composed of a Si tracker, containing 14 planes 
of Si microstrip detectors (121 $\mu$m pitch), 
interleaved with tungsten layers used as
a photon pair converter (each layer is 0.07 radiation length - $X_{0}$). 
At the bottom of the Si tracker a mini-calorimeter - two planes of CsI
bars, for a total on-axis radiation length of 1.5 $X_{0}$ - is 
in charge of the total absorption of the created pairs. The same 
detector, however, can be used for independent detection and
triggering of high energy (300 keV - 200 MeV) transient events.
The hard X-ray section, SuperAGILE, is located on top of the Si-tracker. 
All the above parts are surrounded by an
anticoincidence made of a 6-mm thick (5-mm on the top shield), 
segmented plastic scintillator.

\section{The SuperAGILE Assembly}

SuperAGILE is basically composed of a Detection Plane (DP), a
Collimator equipped with a Coded Mask, a Front-End Electronics and an
Interface Electronics (SAIE). Table 1 resumes the main SuperAGILE
characteristics and Figure 1 shows its appearance
(see also Soffitta et al. 2000 for a more extensive description).

The DP is composed of 4 detection units (DUs), placed on a single Al
honeycomb plane support, so that two of them sample the X-direction and
the other two are devoted to the Y-direction. Each DU is composed of 4 Si
microstrip tiles, bonded in pairs so that the effective length of each
strip is approximately 19 cm. They are read-out through a set of 
IDE AS-XAA1 chips, based on ASIC technology, 12 for each of the DUs.
The collimator (present baseline:
500 $\mu$m thick Carbon Fiber field separators, coated with
a 75 $\mu$m Gold layer) is mounted on the same tray supporting the DP, and in
turn supports the 4 orthogonal, one-dimensional coded masks.
The coded masks have been designed based on an Hadamard sequence,
with a 50\% covering factor. They will be manufactured either in Gold
or in Tungsten, 100 $\mu$m thick.
The SAIE is in charge of interfacing SuperAGILE with the AGILE
Data Handling System, allowing an event-by-event transmission 
with better than 5 $\mu$s timing resolution.
The energy information will be provided in the extended energy range 
between 1 and 64 keV, with 64 channels, to allow for a finer 
threshold calibration at low energies, and exploit for calibration
purposes the Tungsten fluorescences at $\sim$58~keV. 

The combined capabilities of the SAIE and the AGILE Data Handling (see also
Morselli et al. 2000) will allow the transmission to ground of a
relatively large set of scientific housekeeping data, including ratemeters
and detector images. In particular, the AGILE Data Handling will be
able to perform a continuous automatic search for transient events
(e.g., gamma-ray bursts) on timescales from 1~ms to 100~s. 
Once a transient event is triggered onboard, the Data Handling will
be able to provide attitude-corrected sky images for it, determining
the location of the transient source on the sky.
The possibility to distribute in almost real time the coordinates 
of a transient event through a fast link (e.g., TDRSS or similar) 
is currently under study.

\begin{table}
\label{tab}
\begin{center}
\caption{Characteristics of SuperAGILE}
\begin{tabular}{|l|l|}
 \hline
Detector Type            & Silicon Strip             \\ \hline
Basic Detection Unit     & 4 Si tiles,               \\
                         & 19 cm x 19 cm             \\ \hline
Total geometric Area     & 1444 cm$^{2}$             \\ \hline
Nominal Energy Range     & 10-40~keV                 \\ \hline
On-axis Effective Area   & 320 cm$^{2}$ (13 keV)     \\ \hline
Detector Strip Size      & 121 $\mu$m                \\ \hline
Detector Thickness       & 410 $\mu$m                \\ \hline
Energy Resolution (FWHM) & $\sim$3-4 keV             \\ \hline
Timing Accuracy          & $\sim$5 $\mu$s            \\ \hline
Collimator Materials     & 75 $\mu$m Gold-coated     \\
                         & Carbon Fiber              \\ \hline
Mask Size	         & 1444 cm$^{2}$             \\ \hline
Mask-Detector Distance   & 14 cm                     \\ \hline
Mask Transparency	 & 50\%                      \\ \hline
Mask Material            & Gold or Tungsten          \\ \hline
Mask Thickness           & 100 $\mu$m                \\ \hline
Mask Element Size        & 242 $\mu$m                \\ \hline  
Field of View (FWZR)     & 107$^{\circ}$ x 68$^{\circ}$ \\ \hline
On-axis Angular Resolution & 5.9 arcmin \\ \hline
Source Location Accuracy   & $\sim$1-2 arcmin        \\
                           & for bright sources      \\ \hline
\end{tabular}
\end{center}
\end{table}

\begin{figure}
\epsfxsize=13cm
\epsfxsize=8cm
\epsfbox{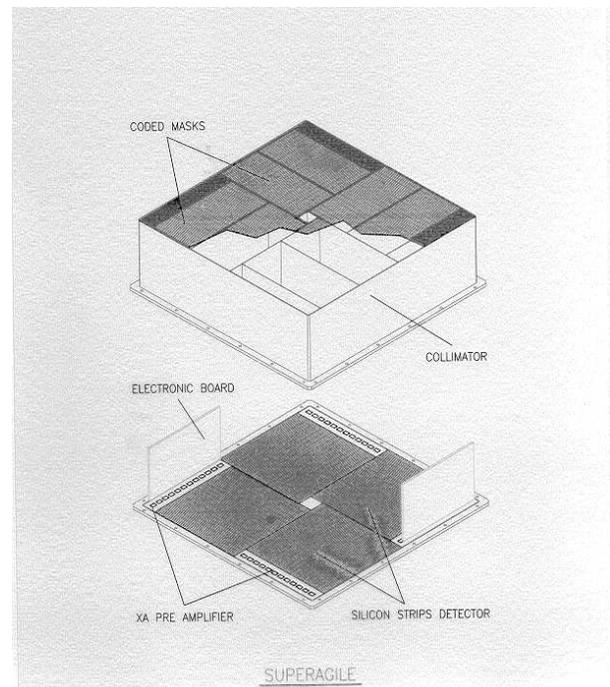}
\caption{Schematic view of the SuperAGILE structure}
\end{figure}

\begin{figure}
\epsfxsize=8cm
\epsfxsize=8cm
\epsfbox{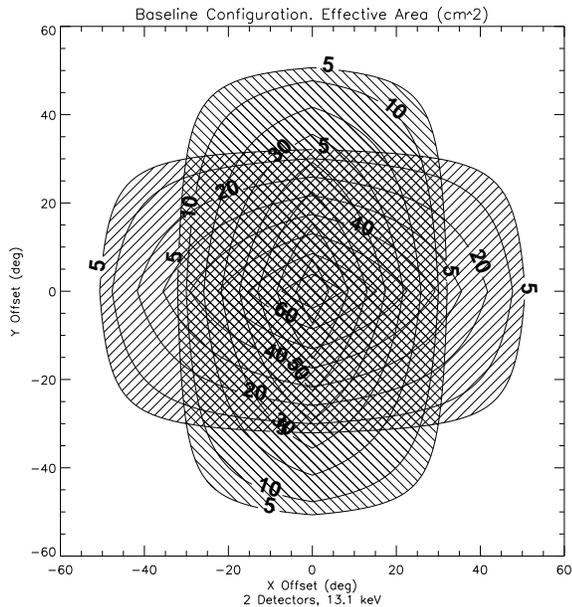}
\caption{Effective area at 13 keV of 2 crossed SuperAGILE detector units.
Numbers give the area in cm$^{2}$ for the individual detector.}
\end{figure}

\section{Laboratory Tests on Prototypes}

The most critical items for the SuperAGILE design were the signal-to-noise
ratio at low energies (i.e., around 10 keV) and the power consumption. 
For this reason extensive
laboratory tests have been performed on the ASIC chips planned to be
used as
front-end electronics: the XA1.3 chip, precursor of the XAA1 chip,
especially developed by IDE AS for SuperAGILE. 
The results of these tests (see  
Del Monte et al. 2000 for a detailed discussion) 
show that the electronic noise can be reduced to 
less than 4 keV (FWHM) and the power consumption to less than
1 mW per channel (SuperAGILE includes 6144 independent electronic
channels). The same tests have shown a critical dependence of 
several chip characteristics (gain, offset, strip address and others) 
from the temperature.

\section{Sensitivity}

We studied the sensitivity and expected astrophysical performances by
means of analytical calculations and Monte Carlo simulations.
In Figure 2 we present the map of the SuperAGILE effective area for
two individual DUs over their field of view (FOV), for a monochromatic 
13.1 keV photon beam (this energy corresponds to the peak in the
area vs. energy relation). The area of two
orthogonal units is presented, showing the overlap of their FOV, providing
an effective bi-dimensional source location capability within the central 
60$^{\circ}$ x 60$^{\circ}$ part of the FOV, in addition to the 
one-dimensional location capability for the further $\sim$20$^{\circ}$. 
The maximum effective area
at the center of the FOV is $\sim$80~cm$^{2}$, thus giving a maximum 
of 320~cm$^{2}$ when the four units are considered together.

In Figure 3 the 5-$\sigma$ sensitivity is presented for sources with 
Crab-like energy spectra, for an integration time of 50 ks, 
over the full 10-40~keV energy range. 
The sensitivity is presented as a function of the source
location within the field of view, along the two directions. 
It is worth noticing the wide central part of the FOV with 
slowly variable sensitivity, allowing for an excellent use 
of most of the FOV for source monitoring purposes.

\begin{figure}
\epsfxsize=8cm
\epsfxsize=8cm
\epsfbox{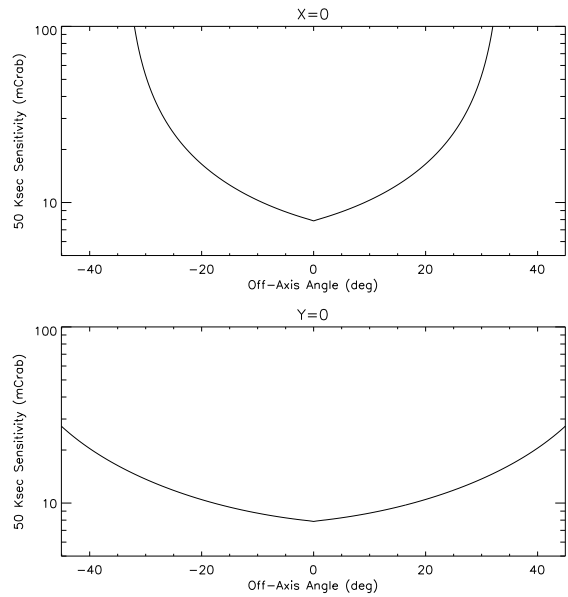}
\caption{50 ks sensitivity of one SuperAGILE detector unit, as a function
of the off-axis angle, in the energy range 10-40~keV.}
\end{figure}

\section{Expected Astrophysical Performances}

In this section we present the expected performances of SuperAGILE
for few interesting classes of hard X-ray sources. In Figure 4 
we show the sensitivity (5-$\sigma$ in 50 ks) as a function of energy.
On the same plot we show the typical 
energy spectrum of the X-ray binary pulsar Her X-1, as observed by 
BeppoSAX near
the maximum of the 35-day cycle (Dal Fiume et al., 1998),
showing that SuperAGILE will be able to provide accurate energy 
spectra for this and similar sources, also when it goes to its periodic 
minimum (a factor $\sim$3 fainter).  
As a result of the long pointings ($\sim$2 weeks) driven by the gamma-ray
tracker, SuperAGILE will point the same source(s) for the same long
time, thus providing both accurate energy
spectra over such a long integration time as well as a
monitoring of the fluxes and energy spectra over much shorter timescales. 
In Figure~4 we also show the observed energy spectra (in a flare
and in a dip state) of the Galactic microquasar GRS 1915+105
(Feroci et al. 1999). Also in this case the SuperAGILE sensitivity is
perfectly suited to allow both spectral and temporal variability 
studies. It is worth noticing that the galactic microquasars are
among the primary targets also for the gamma-ray tracker and will
likely be pointed several times over the AGILE lifetime.

\begin{figure}
\epsfxsize=8cm
\epsfxsize=8cm
\epsfbox{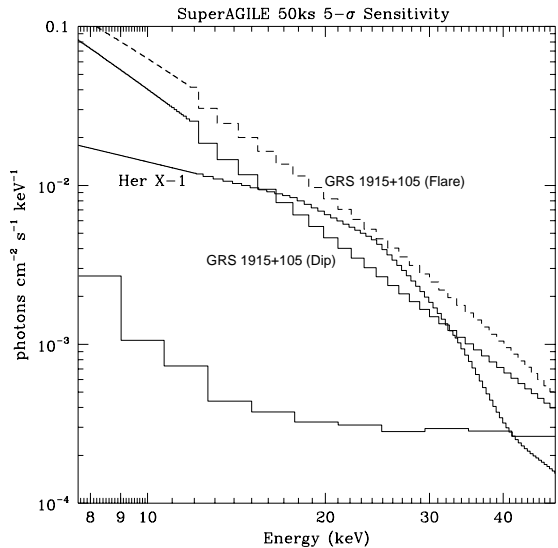}
\caption{On-axis sensitivity of SuperAGILE (5-$\sigma$ in 50 ks) 
compared to the typical energy spectra of the two Galactic sources Her X-1
(from Dal Fiume et al. 1998) and GRS 1915+105 (from Feroci et al. 1999).
}
\end{figure}

In Figure 5 we show the SuperAGILE capability in studying fast hard
X-ray transients, as the short recurrent bursts from the Soft Gamma-ray
Repeaters (e.g., Aptekar et al. 2000). 
The plot clearly shows that also in 250 ms SuperAGILE can 
provide detailed energy spectra of such events. 
The wide field of view of SuperAGILE is very well suited
for a {\it monitoring} of the activity of these sources, that are
mostly concentrated towards the galactic center, thus allowing
to trigger pointed observations when they undergo periods of
intense bursting activity. Furthermore, the giant flares from these
sources (e.g., Feroci et al. 2000) are very good candidates for
emission of rapid and intense flashes of gamma-rays and SuperAGILE
can provide the Tracker with an accurate position of a possible
new soft gamma-ray repeater manifesting itself with a giant flare,
as the source of the 1979 March 5$^{th}$ event did. 

\begin{figure}
\epsfxsize=8cm
\epsfxsize=8cm
\epsfbox{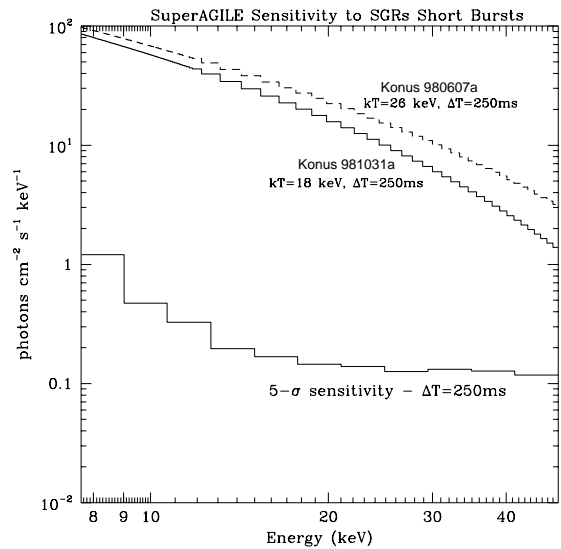}
\caption{
On-axis sensitivity of SuperAGILE (5-$\sigma$ in 250 ms)
compared to two short bursts from the soft gamma-ray repeater 
SGR 1900+14 (Aptekar et al. 2000)
}
\end{figure}

Finally, Figure 6 shows the source location capability
of SuperAGILE for gamma-ray bursts. The plot simulates the sky image
of the SuperAGILE detection of GRB 980425, a relatively weak and
soft gamma-ray burst in the BeppoSAX sample. Although the simulation
assumes an event at 15$^{\circ}$ off-axis in both X and Y directions,
the detection is highly significant and allows for a very good
source location determination, as can be seen by the right-hand panels
where the sky image is zoomed-in.  
This capability motivated the set-up of the onboard 
triggering and source localization system for SuperAGILE (see Sec. 2).

\begin{figure}
\epsfxsize=8cm
\epsfxsize=8cm
\epsfbox{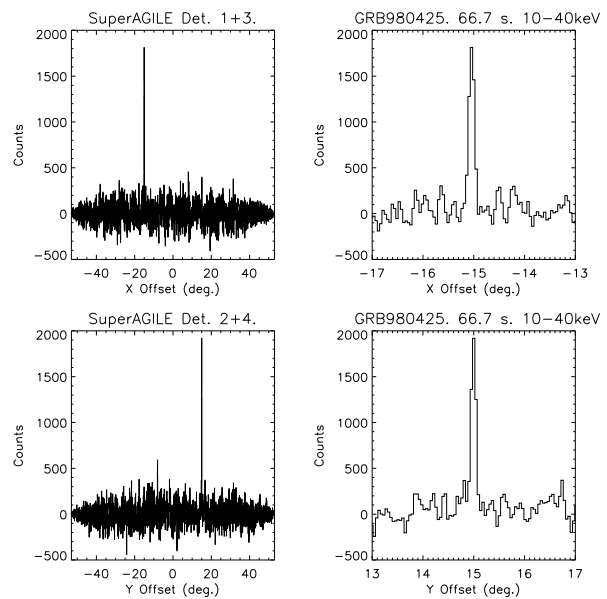}
\caption{
Simulation of the detection  of the relatively weak and soft gamma-ray burst
GRB 980425 at 15$^{\circ}$ off-set in both X and Y directions. 
The left panels show how the event is detected in the sky image in the two
orthogonal directions. The right panels show an enlargement of the same
images. 
}
\end{figure}


\begin{thebibliography}{}
\bibitem[]{}Aptekar, R., et al. 2000, astro-ph/0004402
\bibitem[]{}Barbiellini, G., et al. 2000, Proc. Fifth Compton Symposium,
AIP 510, p.750
\bibitem[]{}Dal Fiume, D., et al. 1998, A\&A 329, L41
\bibitem[]{}Del Monte, E., et al. 2000, Proc. SPIE conference 4140
\bibitem[]{}Feroci, M., et al. 1999, A\&A 351, 985
\bibitem[]{}Feroci, M., et al. 2000, ApJ in press (astro-ph/0010494)
\bibitem[]{}Mereghetti, S., et al. 2000, These proceedings 
\bibitem[]{}Morselli, A., et al. 2000, Proc. SPIE conference 4140
\bibitem[]{}Soffitta, P., et al. 2000, Proc. SPIE conference 4140
\bibitem[]{}Tavani, M., et al. 2000, Proc. Fifth Compton Symposium, 
AIP 510, p.746


\end{thebibliography}
\end{document}